\documentclass[aps,prl,twocolumn,showpacs,superscriptaddress,groupedaddress]{revtex4}  
\usepackage{graphicx}  
\usepackage{graphics}
\usepackage{dcolumn}   
\usepackage{bm}        
\usepackage{amssymb}   

\begin{document}

\title{A Precision Measurement of the Neutron Twist-3 Matrix Element $d_2^n$: \\Probing Color Forces}

\author{M.~Posik}
\email{posik@temple.edu}
\affiliation{Temple University, Philadelphia, PA 19122}

\author{D.~Flay}
\affiliation{Temple University, Philadelphia, PA 19122}

\author{D.~S.~Parno}
\affiliation{Carnegie Mellon University, Pittsburgh, PA 15213}
\affiliation{Center for Experimental Nuclear Physics and Astrophysics, University of Washington, Seattle, WA 98195}

\author{K.~Allada}
\affiliation{University of Kentucky, Lexington, KY 40506}

\author{W.~Armstrong}
\affiliation{Temple University, Philadelphia, PA 19122}

\author{T.~Averett}
\affiliation{College of William and Mary, Williamsburg, VA 23187}

\author{F.~Benmokhtar}
\affiliation{Duquesne University, Pittsburgh, PA 15282}
\affiliation{Carnegie Mellon University, Pittsburgh, PA 15213}

\author{W.~Bertozzi}
\affiliation{Massachusetts Institute of Technology, Cambridge, MA 02139}

\author{A.~Camsonne}
\affiliation{Thomas Jefferson National Accelerator Facility, Newport News, VA 23606}

\author{M.~Canan}
\affiliation{Old Dominion University, Norfolk, VA 23529}

\author{G.D.~Cates}
\affiliation{University of Virginia, Charlottesville, VA 22904}

\author{C.~Chen}
\affiliation{Hampton University, Hampton, VA 23187}

\author{J.-P.~Chen}
\affiliation{Thomas Jefferson National Accelerator Facility, Newport News, VA 23606}

\author{S.~Choi}
\affiliation{Seoul National University, Seoul 151-742, South Korea}

\author{E.~Chudakov}
\affiliation{Thomas Jefferson National Accelerator Facility, Newport News, VA 23606}

\author{F.~Cusanno}
\affiliation{INFN, Sezione di Roma, I-00161 Rome, Italy}
\affiliation{Istituto Superiore di Sanit\`a, I-00161 Rome, Italy}

\author{M.~M.~Dalton}
\affiliation{University of Virginia, Charlottesville, VA 22904}

\author{W.~Deconinck}
\affiliation{Massachusetts Institute of Technology, Cambridge, MA 02139}

\author{C.W.~de~Jager}
\affiliation{Thomas Jefferson National Accelerator Facility, Newport News, VA 23606}

\author{X.~Deng}
\affiliation{University of Virginia, Charlottesville, VA 22904}

\author{A.~Deur}
\affiliation{Thomas Jefferson National Accelerator Facility, Newport News, VA 23606}

\author{C.~Dutta}
\affiliation{University of Kentucky, Lexington, KY 40506}

\author{L.~El~Fassi}
\affiliation{Old Dominion University, Norfolk, VA 23529}
\affiliation{Rutgers, The State University of New Jersey, Piscataway, NJ 08855}

\author{G.~B.~Franklin}
\affiliation{Carnegie Mellon University, Pittsburgh, PA 15213}

\author{M.~Friend}
\affiliation{Carnegie Mellon University, Pittsburgh, PA 15213}

\author{H.~Gao}
\affiliation{Duke University, Durham, NC 27708}

\author{F.~Garibaldi}
\affiliation{INFN, Sezione di Roma, I-00161 Rome, Italy}

\author{S.~Gilad}
\affiliation{Massachusetts Institute of Technology, Cambridge, MA 02139}

\author{R.~Gilman}
\affiliation{Thomas Jefferson National Accelerator Facility, Newport News, VA 23606}
\affiliation{Rutgers, The State University of New Jersey, Piscataway, NJ 08855}

\author{O.~Glamazdin}
\affiliation{Kharkov Institute of Physics and Technology, Kharkov 61108, Ukraine}

\author{S.~Golge}
\affiliation{Old Dominion University, Norfolk, VA 23529}

\author{J.~Gomez}
\affiliation{Thomas Jefferson National Accelerator Facility, Newport News, VA 23606}

\author{L.~Guo}
\affiliation{Los Alamos National Laboratory, Los Alamos, NM 87545}

\author{O.~Hansen}
\affiliation{Thomas Jefferson National Accelerator Facility, Newport News, VA 23606}

\author{D.~W.~Higinbotham}
\affiliation{Thomas Jefferson National Accelerator Facility, Newport News, VA 23606}

\author{T.~Holmstrom}
\affiliation{Longwood University, Farmville, VA 23909}

\author{J.~Huang}
\affiliation{Massachusetts Institute of Technology, Cambridge, MA 02139}

\author{C.~Hyde}
\affiliation{Old Dominion University, Norfolk, VA 23529}
\affiliation{Universit\'e Blaise Pascal/IN2P3, F-63177 Aubi\`ere, France}

\author{H.~F.~Ibrahim}
\affiliation{Cairo University, Giza 12613, Egypt}

\author{X.~Jiang}
\affiliation{Rutgers, The State University of New Jersey, Piscataway, NJ 08855}
\affiliation{Los Alamos National Laboratory, Los Alamos, NM 87545}

\author{ G.~Jin}
\affiliation{University of Virginia, Charlottesville, VA 22904}

\author{J.~Katich}
\affiliation{College of William and Mary, Williamsburg, VA 23187}

\author{A.~Kelleher}
\affiliation{College of William and Mary, Williamsburg, VA 23187}

\author{A.~Kolarkar}
\affiliation{University of Kentucky, Lexington, KY 40506}

\author{W.~Korsch}
\affiliation{University of Kentucky, Lexington, KY 40506}

\author{G.~Kumbartzki}
\affiliation{Rutgers, The State University of New Jersey, Piscataway, NJ 08855}

\author{J.J.~LeRose}
\affiliation{Thomas Jefferson National Accelerator Facility, Newport News, VA 23606}

\author{R.~Lindgren}
\affiliation{University of Virginia, Charlottesville, VA 22904}

\author{N.~Liyanage}
\affiliation{University of Virginia, Charlottesville, VA 22904}

\author{E.~Long}
\affiliation{Kent State University, Kent, OH 44242}

\author{A.~Lukhanin}
\affiliation{Temple University, Philadelphia, PA 19122}

\author{V.~Mamyan}
\affiliation{Carnegie Mellon University, Pittsburgh, PA 15213}

\author{D.~McNulty}
\affiliation{University of Massachusetts, Amherst, MA 01003}

\author{Z.-E.~Meziani}
\email{meziani@temple.edu}
\affiliation{Temple University, Philadelphia, PA 19122}

\author{R.~Michaels}
\affiliation{Thomas Jefferson National Accelerator Facility, Newport News, VA 23606}

\author{M.~Mihovilovi\v{c}}
\affiliation{Jo\v{z}ef Stefan Institute, SI-1000 Ljubljana, Slovenia}

\author{B.~Moffit}
\affiliation{Massachusetts Institute of Technology, Cambridge, MA 02139}
\affiliation{Thomas Jefferson National Accelerator Facility, Newport News, VA 23606}

\author{N.~Muangma}
\affiliation{Massachusetts Institute of Technology, Cambridge, MA 02139}

\author{S.~Nanda}
\affiliation{Thomas Jefferson National Accelerator Facility, Newport News, VA 23606}

\author{A.~Narayan}
\affiliation{Mississippi State University, MS 39762}

\author{V.~Nelyubin}
\affiliation{University of Virginia, Charlottesville, VA 22904}

\author{B.~Norum}
\affiliation{University of Virginia, Charlottesville, VA 22904}

\author{Nuruzzaman}
\affiliation{Mississippi State University, MS 39762}

\author{Y.~Oh}
\affiliation{Seoul National University, Seoul 151-742, South Korea}

\author{J.~C.~Peng}
\affiliation{University of Illinois at Urbana-Champaign, Urbana, IL 61801}

\author{X.~Qian} 
\affiliation{Duke University, Durham, NC 27708}
\affiliation{Kellogg Radiation Laboratory, California Institute of Technology, Pasadena, Ca 91125}

\author{Y.~Qiang}
\affiliation{Duke University, Durham, NC 27708}
\affiliation{Thomas Jefferson National Accelerator Facility, Newport News, VA 23606}

\author{A.~Rakhman}
\affiliation{Syracuse University, Syracuse, NY 13244}

\author{S.~Riordan}
\affiliation{University of Virginia, Charlottesville, VA 22904}
\affiliation{ University of Massachusetts, Amherst, MA 01003}

\author{A.~Saha} \thanks{Deceased}
\affiliation{Thomas Jefferson National Accelerator Facility, Newport News, VA 23606}

\author{B.~Sawatzky}
\affiliation{Temple University, Philadelphia, PA 19122}
\affiliation{Thomas Jefferson National Accelerator Facility, Newport News, VA 23606}

\author{M.~H.~Shabestari}
\affiliation{University of Virginia, Charlottesville, VA 22904}

\author{A.~Shahinyan}
\affiliation{Yerevan Physics Institute, Yerevan 375036, Armenia}

\author{S.~\v{S}irca}
\affiliation{University of Ljubljana, SI-1000 Ljubljana, Slovenia}
\affiliation{Jo\v{z}ef Stefan Institute, SI-1000 Ljubljana, Slovenia}

\author{P.~Solvignon}
\affiliation{Argonne National Lab, Argonne, IL 60439}
\affiliation{Thomas Jefferson National Accelerator Facility, Newport News, VA 23606}

\author{R.~Subedi}
\affiliation{University of Virginia, Charlottesville, VA 22904}

\author{V.~Sulkosky}
\affiliation{Massachusetts Institute of Technology, Cambridge, MA 02139}
\affiliation{Thomas Jefferson National Accelerator Facility, Newport News, VA 23606}

\author{W.~A.~Tobias}
\affiliation{University of Virginia, Charlottesville, VA 22904}

\author{W.~Troth}
\affiliation{Longwood University, Farmville, VA 23909}

\author{D.~Wang}
\affiliation{University of Virginia, Charlottesville, VA 22904}

\author{Y.~Wang}
\affiliation{University of Illinois at Urbana-Champaign, Urbana, IL 61801}

\author{B.~Wojtsekhowski}
\affiliation{Thomas Jefferson National Accelerator Facility, Newport News, VA 23606}

\author{X.~Yan}
\affiliation{University of Science and Technology of China, Hefei 230026, People's Republic of China}

\author{H.~Yao}
\affiliation{Temple University, Philadelphia, PA 19122}
\affiliation{College of William and Mary, Williamsburg, VA 23187}

\author{Y.~Ye}
\affiliation{University of Science and Technology of China, Hefei 230026, People's Republic of China}

\author{Z.~Ye}
\affiliation{Hampton University, Hampton, VA 23187}

\author{L.~Yuan}
\affiliation{Hampton University, Hampton, VA 23187}

\author{X.~Zhan}
\affiliation{Massachusetts Institute of Technology, Cambridge, MA 02139}

\author{Y.~Zhang}
\affiliation{Lanzhou University, Lanzhou 730000, Gansu, People's Republic of China}

\author{Y.-W.~Zhang}
\affiliation{Lanzhou University, Lanzhou 730000, Gansu, People's Republic of China}
\affiliation{Rutgers, The State University of New Jersey, Piscataway, NJ 08855}

\author{B.~Zhao}
\affiliation{College of William and Mary, Williamsburg, VA 23187}

\author{X.~Zheng}
\affiliation{University of Virginia, Charlottesville, VA 22904}

\collaboration{The Jefferson Lab Hall A Collaboration}
\noaffiliation



\begin{abstract}
Double-spin asymmetries and absolute cross sections were measured at large Bjorken $x$ (0.25 $ \le x \le $ 0.90), in both the deep-inelastic and resonance regions, by scattering longitudinally polarized electrons at beam energies of 4.7 and 5.9 GeV from a transversely and longitudinally polarized $^3$He target. In this dedicated experiment, the spin structure function $g_2^{^3\text{He}}$ was determined with precision at large $x$, and the neutron twist-three matrix element $d_2^n$ was measured at $\left< Q^2\right>$ of 3.21 and 4.32 GeV$^2$/$c^2$, with an absolute precision of about $10^{-5}$. Our results are found to be in agreement with lattice QCD calculations and resolve the disagreement found with previous data at $\left< Q^2\right> =$ 5 GeV$^2$/$c^2$. Combining $d_2^n$ and a newly extracted twist-four matrix element, $f_2^n$, the average neutron color electric and magnetic forces were extracted and found to be of opposite sign and about 30 MeV/fm in magnitude.\end{abstract}

\pacs{12.38.Aw, 12.38.Qk, 13.88.+e, 14.20.Dh}
\maketitle

 Over the past 30 years, the availability of polarized targets and lepton beams has enabled an intensive worldwide experimental program of inclusive deep-inelastic scattering (DIS) measurements focused on the investigation of the nucleon spin structure~\cite{Aidala:2012mv}.~This led to the confirmation of the Bjorken sum rule~\cite{Bjorken:1966,Bjorken:1970}, a fundamental sum rule of quantum chromodynamics (QCD), and the determination of the quarks' spin contribution to the total nucleon spin~\cite{Aidala:2012mv}.~Exploration of the nucleon spin structure through QCD has shown that both the $g_1$ and $g_2$ spin structure functions of the nucleon contain contributions from the elusive quark-gluon correlations~\cite{Shuryak:1981pi,Jaffe:1989xx,Jaffe:1990qh} beyond the perturbative-QCD radiative corrections~\cite{Altarelli:1977zs,Gribov:1972ri, Dokshitzer:1977}. In the case of $g_1$, these correlations emerge at a higher order in the perturbative expansion in powers of the inverse $Q^2$ ($Q^2$ is defined as $-q^2$, where $q^2$ is the virtual photon's four-momentum transfer squared) and thus are suppressed; however, they contribute at leading order in $g_2$.~These correlations manifest themselves in $\bar{g_2}$, a deviation of the measured $g_2$ from the so-called Wandzura-Wilczek value $g_2^{\text{WW}}$~\cite{Wandzura:1977} that is expressed solely in terms of the $g_1$ spin structure function:
 \begin{eqnarray}
 \bar g_2(x,Q^2) &=& g_2(x,Q^2) - g_2^{\text{WW}}(x,Q^2), \\
g_2^{\text{WW}}(x,Q^2) &=& -g_1(x,Q^2) + \int_x^1 g_1(y,Q^2) dy/y,
 \end{eqnarray}
where $x$ is the Bjorken variable interpreted in the infinite momentum frame as being the fraction of the nucleon's longitudinal momentum carried by the leading struck quark in the DIS process.~At present there are no {\it ab initio} calculations of $\bar{g_2}$. Nevertheless using the operator product expansion~\cite{Shuryak:1981pi,Jaffe:1990qh}, the $Q^2$-dependent quantity  
\begin{eqnarray}
d_2&=&3\int_0^1 dx x^2  \bar g_2(x) = \int_0^1 dx x^2 \left [ 3g_2(x) + 2 g_1(x) \right ]
\end{eqnarray}
\noindent{}can be related to a specific matrix element containing local operators of quark and gluon fields~\cite{Ji:1993sv,Stein:1995si}, and is calculable in lattice QCD~\cite{Gockeler:2005vw}.~Insight into the physical meaning of $d_2$ was articulated by Ji~\cite{Ji:1995qe} who expressed $d_2$ in terms of a linear combination of $\chi_E$ and $\chi_B$, dubbed the electric and magnetic ``color polarizabilities,'' summed over the quark flavors:
\begin{eqnarray}
\chi_E \vec S &=& \frac{1}{2M^2}\langle P,S | q^{\dagger} \vec \alpha \times g \vec E q | P,S \rangle , \\
\chi_B \vec S &=& \frac{1}{2M^2}\langle P,S | q^{\dagger} g \vec B q | P,S \rangle, 
\end{eqnarray}
where $P$ and $S$ are the nucleon momentum and spin, $q$ and $q^{\dagger}$ the quark fields, $\vec E$ and $\vec B$ the average color electric and magnetic fields seen by the struck quark, $\vec \alpha$ the velocity of the struck quark, $g$ the strong coupling parameter, and $M$ the nucleon mass. 

More recently, Burkardt~\cite{Burkardt:2013} identified $d_2$ as being proportional to the instantaneous average sum of the transverse electric $F_E$ and magnetic $F_B$ color forces the struck quark experiences at the instant it is hit by the virtual photon due to the remnant di-quark system in the DIS process. The net average force contributes to what is called the ``chromodynamic lensing'' effect in semi-inclusive DIS where the struck quark experiences
color forces as it exits the nucleon before converting into an outbound hadron~\cite{Burkardt:2013,Burkardt:2003uw}. The relations between the color forces, the color polarizabilities and the matrix elements of the quark-gluon correlations are given by:
\begin{eqnarray}
\label{cf:f2d2_1}
F_E &=& -\frac{M^2}{4}\chi_E = -\frac{M^2}{4}\left [ \frac{2}{3} \left ( 2d_2 + f_2 \right ) \right ], \\
F_B &=& -\frac{M^2}{2}\chi_B =  -\frac{M^2}{2}\left [ \frac{1}{3} \left ( 4d_2 - f_2 \right ) \right ],
\label{cf:f2d2}
\end{eqnarray}
\noindent{}where $f_2$ is a twist-4 matrix element of the quark-gluon correlation.~This twist-4 matrix element cannot be measured directly, but may be extracted by taking advantage of the $Q^2$ dependence of $g_1(x,Q^2)$~\cite{Meziani:2004ne,Osipenko:2004xg}, since it appears as a higher-order contribution suppressed by $1/Q^2$ in $\Gamma_{1}$, the first moment of $g_1$.

The neutron $d_2$ ($d_2^n$) has been calculated in different nucleon structure models~\cite{Stein:1994zk,Balitsky:1989jb,Song:1996ea,Weigel:1996jh} and in lattice QCD~\cite{Gockeler:2005vw}. The results consistently give a small negative value deviating from the measured positive value by about 2 standard deviations~\cite{Anthony:2002hy,Zheng:2004ce}.~At a fundamental level, the forces $d_2$ probes are in part responsible for the confinement of the constituents; determining and understanding them in QCD is an important goal. The disagreement between the experimental and theoretical results of $d_2^n$ called for a dedicated measurement.

The E06-014 experiment~\cite{Choi:2006} was performed at Jefferson Lab (JLab) in Hall A~\cite{Alcorn:2004sb} from February to March of 2009. In this experiment measurements of inclusive scattering of a longitudinally polarized electron beam, at an average current of 15 $\mu A$, from a polarized $^3$He target were carried out at two incident beam energies of 4.7 GeV and 5.9 GeV and with two states of target polarization, transverse (perpendicular to the electron beam in the scattering plane) and longitudinal (along the electron beam).~Scattered electrons with momenta ranging from 0.7 to 2 GeV/c were detected in both the BigBite spectrometer (BBS)~\cite{deLange:1998qq} and the left high-resolution spectrometer (LHRS)~\cite{Alcorn:2004sb} each set at a scattering angle of 45$^{\circ}$. The large momentum and angular acceptance ($\sim$ 64 msr) of the BBS allowed it to precisely measure the double-spin asymmetries (DSA) over the full momentum range at one  current setting of the spectrometer. The absolute cross sections were obtained from the well understood LHRS by scanning over the same momentum range in discrete steps. The longitudinal (transverse) DSA is defined as
\begin{equation}
A_{\parallel(\perp)} = \frac{1}{P_t P_bD_{\text{N}_2} }\frac{1}{(\cos\phi)}\frac{N^{\downarrow\Uparrow(\Rightarrow)}-N^{\uparrow\Uparrow(\Rightarrow)}}{N^{\downarrow\Uparrow(\Rightarrow)}+N^{\uparrow\Uparrow(\Rightarrow)}},
\label{eq:Apar}
\end{equation}
\noindent{}where $N$ is the number of detected electrons, $P_t$ is the target polarization, $P_b$ is the electron beam polarization, $D_{\text{N}_2}$ is the nitrogen dilution factor that accounts for the small amount ($\sim$ 1\% of $^3$He density) of N$_2$ present in the $^{3}$He target to reduce depolarization effects, and $\phi$ is the angle between the scattering plane, defined by the initial and final electron momenta, and the polarization plane, defined by the electron and target spins (with $\cos\phi$ applied only to the transverse asymmetry). The single arrows refer to the electron helicity direction and the double arrows refer to the target spin direction. The orientation of the latter are such that arrows pointing to the right (left) represent spins that are transverse to the electron momentum pointing towards the BBS (LHRS), while arrows pointing up (down) represent spins parallel (anti-parallel) to the electron momentum.

The incident electron beam polarization was measured using two independent polarimeters based on M\o ller~\cite{Glamazdin:1999gg} and Compton~\cite{Escoffier:2005gi,Friend:2011qh} scattering, whose combined analysis for three run periods yielded electron beam polarizations of 74\% $\pm$ 1\% ($E$ = 5.9 GeV), 79\% $\pm$ 1\% ($E$ = 5.9 GeV) and 63\% $\pm$ 1\% ($E$ = 4.7 GeV)~\cite{parno_thesis}.~The residual beam-charge asymmetry was controlled to within 100 ppm through the use of a feedback loop.
In a polarized $^3$He target, the neutron carries the greater part of the nuclear spin, which allows it to serve as an effective neutron target~\cite{Bissey:2001cw}.~The scattered electrons interacted with about 10.5 amg (in-beam conditions) of polarized $^3$He gas contained in a 40-cm-long target cell. The $^{3}$He nuclei were polarized via double spin-exchange optical pumping of a Rb-K mixture~\cite{Babcock:2003zz}.~Nuclear magnetic resonance (NMR) measurements were taken approximately every 4 hours to monitor the target polarization.~The relative NMR measurements were calibrated with absolute electron paramagnetic resonance measurements, which were taken every few days.~Additionally, the longitudinal target polarization was cross-checked using water NMR measurements.~The average target polarization achieved was 50.5\% $\pm$ 3.6\%~\cite{posik_thesis}.\\
\indent{}The BBS consisted of a large-aperture dipole magnet in front of a detector stack whose configuration was similar to that used in~\cite{Qian:2011py} except for the addition of a newly built threshold gas \v{C}erenkov detector~\cite{posik_thesis} used for both electron and positron identification and pion rejection.~The magnetic optics software package used for the BBS was calibrated at an incident energy of 1.2 GeV using various targets described in~\cite{Qian:2011py}.~Angular and momentum resolutions of 10 mrad and 1\% were achieved, respectively~\cite{parno_thesis,posik_thesis}.~To keep trigger rates compatible with a high data-acquisition live time ($\gtrsim 80$\%), the main electron trigger was formed when signals above threshold were registered in geometrically overlapping regions of the calorimeter and gas \v{C}erenkov. The BBS achieved a total pion rejection factor of better than 10$^4$.\\ 
\indent{}The LHRS is a small-acceptance spectrometer ($\sim$ 6~msr) and was used with its standard detector package~\cite{Alcorn:2004sb}. Optics calibrations~\cite{Qian:2011py} for the LHRS used the same targets that were used to calibrate the BBS optics. The LHRS achieved a pion rejection factor of better than 10$^{5}$, and measured the $e^-$--$^3$He inclusive cross section to better than 8\% (includes statistical and systematic uncertainties added in quadrature).\\
\indent{}The two possible sources of background contamination of the electron sample were charged pions and pair-produced electrons resulting from the conversion of $\pi^0$ decay photons.~The BBS $\pi^{-}$~($\pi^{+}$) contamination of the $e^-$~($e^+$) sample was estimated from the pre-shower energy spectrum and found to be less than 3\% (6.5\%) across the momentum acceptance for $\pi^-$~($\pi^+$). Weighting the measured $\pi^{\pm}$ asymmetries by the pion contamination resulted in a negligible pion asymmetry contamination to the electron DSA. The $\pi^{\pm}$ contamination measured in the LHRS was negligible.\\
\indent{}
We measured the electron background by assuming it was due to symmetric pair-produced $e^+e^-$ events. Reversing the BBS and LHRS magnet polarities results in positrons rather than electrons being steered into the detectors.~By switching the magnet polarity both electrons and positrons should see the same acceptance which then drops out when forming the $e^+/e^-$ ratio. The positron cross section was measured with the LHRS and subtracted from the electron cross section.~We used the BBS measured and fitted $e^+/e^-$ ratios, along with statistically-weighted positron asymmetry measurements, to determine the amount of pair-production contamination in the electron sample. The $e^+/e^-$ ratio at low $x$ ($\left<x\right>$ = 0.277) was about 56\%, and quickly fell off to below 1\% ($\left<x\right>$ = 0.673) as $x$ increased. The positron asymmetry was measured with the BBS magnet in normal polarity to be about 1-2\%. The positron asymmetries were cross-checked by reversing the BBS magnet polarity and measuring the positron asymmetry for one DSA setting. The background and false asymmetries were removed from the electron asymmetries according to: 

\begin{equation}  \label{eq:Acor}
   A^{e^-}_{\perp,\parallel} = \frac{A^{m,e^-}_{\perp,\parallel}-c_3A_{\perp,\parallel}^{m,e^+}}{1-c_1-c_3+c_2c_3}, \\
\end{equation}   

\noindent where $c_1$ is the $\pi^-/e^-$ ratio; $c_2$ is the $\pi^+/e^+$ ratio; 
$c_3$ is the $e^+/e^-$ ratio and $A^{m,e^-(e^+)}_{\perp,\parallel}$ is the measured 
electron (positron) asymmetry.

\begin{figure}[b!]
\includegraphics[width=\columnwidth, angle=0]{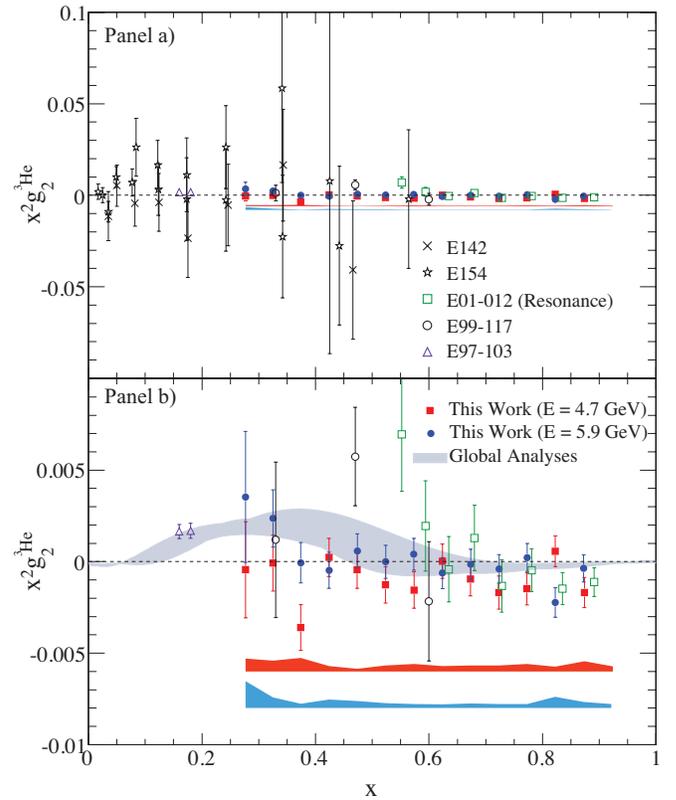}
\caption{(Color) $x^2$-weighted $g_2^{^3\text{He}}$ plotted against $x$ for data from E99-117~\cite{Zheng:2004ce}, E97-103~\cite{Kramer:2005qe}, E142~\cite{Anthony:1996mw}, E154~\cite{Abe:1997qk}, and E01-012~\cite{Solvignon:2013yun} with $Q^2 > 1$ GeV$^2$/c$^2$. Panel a): All error bars on the world data represent statistical and systematic uncertainties added in quadrature. Panel b): The error bars on our data are statistical only.~The top (red) and bottom (blue) bands represent the systematic uncertainty associated with the E = 4.7 and 5.9 GeV data sets, respectively. The grey band shows the $g_2^{\text{WW},^3\text{He}}$ coverage from several global analyses~\cite{deFlorian:2008mr,Bourrely:2001du,Bourrely:2007if,deFlorian:2005mw,Leader:2005ci,Gehrmann:1995ag}.}
\label{fig:g2}
\end{figure}

Lastly, the electromagnetic internal and external radiative corrections were performed on the unpolarized cross section $\sigma_0$ 
using the formalism of Mo and Tsai~\cite{MoTsai:1969,Tsai:1971qi}.~The elastic~\cite{MoTsai:1969} and quasi-elastic~\cite{Stein:1975yy} 
radiative tails were subtracted using form factors from~\cite{Amroun:1994qj} and~\cite{Lightbody:1988}.~The inelastic corrections were evaluated using 
the F1F209 parameterization~\cite{Bosted:2012qc} for the unmeasured cross sections in the resonance and DIS regions.~We followed the formalism of Akushevich {\it et  al.}~\cite{Akushevich:1997di} to perform the radiative corrections on $\Delta\sigma_{\parallel,\perp} = 2\sigma_0A_{\parallel,\perp}$, the polarized 
cross-section differences. Here, the tails from the polarized elastic cross-section differences were found to be negligible. The remaining quasi-elastic~\cite{Bosted:1994tm,Amaro:2004bs}, resonance~\cite{Drechsel:2007if}, and deep-inelastic regions~\cite{deFlorian:2008mr} were treated together using inputs from their respective models.~The size of the total correction in all cases did not exceed 45\% of the measured $\sigma_0$, $\Delta\sigma_{\parallel}$, and $\Delta\sigma_{\perp}$. Although the magnitude of this correction is significant, the associated absolute uncertainty on the radiative corrections on $g_1$ and $g_2$ were less than 5\%, which is smaller than their statistical uncertainty.\\
\indent{}In Fig.~\ref{fig:g2} we show the $x^2$-weighted polarized spin structure function $g_2$ of $^3$He, formed from the measured Born asymmetries and cross sections according to 
\begin{eqnarray}
g_2^{^3\text{He}} &=& \frac{MQ^2}{4\alpha^2} \frac{y^2\sigma_0}{\left( 1-y \right)\left( 2-y \right)}\times \\
           & &\left[-A^{e^-}_{\parallel} + \frac{1+\left(1-y\right)\cos\theta}{\left(1-y\right)\sin\theta}A^{e^-}_{\perp}  \right], \nonumber
\label{eq:g2}
\end{eqnarray} 
\noindent{}where $\alpha$ is the electromagnetic coupling constant, $y=\left(E-E'\right)/E$ is the fraction of the incident electron energy loss in the nucleon rest frame, $E$ is the incident electron energy, $E'$ is the scattered electron energy, and $\theta$ is the electron scattering angle.~Note the dramatic improvement of the statistical precision of our data in Fig.~1a compared to the available $^3$He world data; Fig.~1b is zoomed in by a factor of 10 and shows only a subset of the world data.

The measured DSAs and cross sections at each beam energy were used to evaluate $d_2^{^3\text{He}}$ at two $\left<Q^2\right>$ values (3.21 and 4.32 GeV$^2$/$c^2$) according to
\begin{eqnarray}
\label{eq:d2}
d_2^{^3\text{He}} &=& \displaystyle \int_{0.25}^{0.90} dx\frac{MQ^2}{4\alpha^2} \frac{x^2y^2\sigma_0}{\left(1-y\right)\left(2-y\right)} \times\\ 
           &&\hspace{-1cm} \left[ \left( 3\frac{1+\left(1-y\right)\cos\theta}{\left(1-y\right)\sin\theta} + \frac{4}{y}\tan\frac{\theta}{2} \right)A^{e^-}_\perp + \left( \frac{4}{y} - 3 \right)A^{e^-}_\parallel \right].\nonumber
\end{eqnarray}
\noindent{}The upper integration limit of $x$ = 0.90 was chosen in order to avoid the quasi-elastic peak and the $\Delta$ resonance. In addition to using Eq.~\ref{eq:d2}, the Nachtmann moments~\cite{Nachtmann:1973aa} may be used to evaluate $d_2^{^3\text{He}}$, but the difference between the two approaches at our kinematics is smaller than the statistical precision of our measured $d_2^{^3\text{He}}$ value. Neutron information was extracted from $d_2^{^3\text{He}}$ through the expression
\begin{equation}
 d_2^n = \frac{d_2^{^3\text{He}} - \left(2P_p - 0.014\right)d_2^p}{P_n + 0.056},
\label{eq:d2n}
\end{equation}
\noindent{}where $P_p$ and $P_n$ are the effective proton and neutron polarizations in $^3$He, and the factors 0.056 and 0.014 are due to the $\Delta$-isobar contributions~\cite{Bissey:2001cw}. $d_2^p$ in Eq.~\ref{eq:d2n} was calculated from various global analyses~\cite{deFlorian:2008mr,Bourrely:2001du,Bourrely:2007if,deFlorian:2005mw,Leader:2005ci,Gehrmann:1995ag} to be (-17.5$\pm$5.3)$\times10^{-4}$ and (-16.9$\pm$4.7)$\times10^{-4}$ at the kinematics of E06-014 at average $\left<Q^2\right>$ values of 3.21~(where $Q^2$ ranged from about 2.0 to 4.9 GeV$^2$/$c^2$) and 4.32~GeV$^2$/$c^2$~(where $Q^2$ ranged from about 2.6 to 6.6 GeV$^2$/$c^2$), respectively.~Additionally, other neutron extraction methods were studied in~\cite{Ethier:2013hna}; those results were found to be consistent within our total uncertainty. 

\indent{}The $d_2^n$ values measured during E06-014 represent only partial integrals. The full integrals can be evaluated by computing the low- and high-$x$ contributions. The low-$x$ contribution is suppressed due to the $x^2$-weighting of the $d_2$ integrand, and was calculated by fitting existing $g_1^n$~\cite{Anthony:1996mw,Abe:1998wq,Abe:1997qk,Kramer:2005qe} 
and $g_2^n$~\cite{Anthony:1999py,Anthony:2002hy,Kramer:2005qe} data. 
The fits to both structure functions were dominated by the precision data from~\cite{Kramer:2005qe}, and extended in $x$ from 0.02 to 0.25. A possible $Q^2$-dependence of this low-$x$ contribution was presumed to be negligible in this analysis.~The high-$x$ contribution, dominated by the elastic $x$=1 contribution with a negligible contribution from 0.9 $< x <$ 1, was estimated using the elastic form factors $G_E^n$ and $G_M^n$, computed from the parameterizations given in~\cite{Riordan:2010id} and~\cite{Kelly:2004hm}, respectively. The individual contributions used to evaluate the full $d_2^n$ integral are listed in Table~\ref{tbl:d2n}. 

\begin{figure}[!]
\center
\includegraphics[width = \columnwidth]{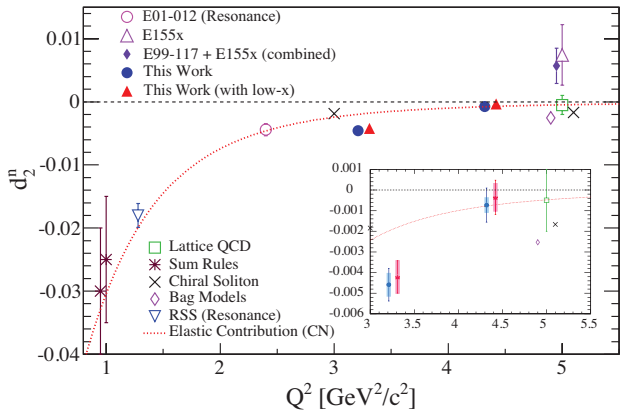}
\caption{(Color) $d_2^n$ data plotted against $Q^2$ for data with $\left< Q^2\right> \ge 1$ GeV$^2$/c$^2$. The error bars on the world data from E01-012~\cite{Solvignon:2013yun}, E155x~\cite{Anthony:2002hy}, E99-117+E155x~\cite{Zheng:2004ce}, and RSS~\cite{Slifer:2008xu} represent the in quadrature sum of their statistical and systematic uncertainties. Our results are displayed with and without the low-$x$ contribution added, and are offset in $Q^2$ for clarity. The inset figure zooms in around our results, with the shaded boxes representing our systematic uncertainty.} 
\label{fig:final_d2n}
\end{figure}

\indent{}The fully integrated $d_2^n$ results from this experiment are shown as a function of $Q^2$ in Fig.~\ref{fig:final_d2n} along with the world data and available calculations.~Our $d_2^n$ results are in agreement with the lattice QCD~\cite{Gockeler:2005vw} (evaluated at $Q^2$ = 5 GeV$^2$/c$^2$), bag model~\cite{Song:1996ea} (evaluated at $Q^2$ = 5 GeV$^2$/c$^2$) and chiral soliton model~\cite{Weigel:1996jh} (evaluated at $Q^2$ = 3 and 5 GeV$^2$/c$^2$) calculations, which predict a small negative value of $d_2^n$ at large $Q^2$. We note that at lower $Q^2$, the elastic contribution of $d_2^n$ dominates the measured values and is in agreement with the QCD sum rule calculations~\cite{Stein:1994zk,Balitsky:1989jb} (evaluated at $Q^2$ = 1 GeV$^2$/c$^2$).~Given our precision, we find a $d_2^n$ value near $Q^2$ = 5 GeV$^2$ that is about 3 standard deviations smaller than the lowest error bar reported by SLAC E155x. 

\begin{table*}[!]
\begin{ruledtabular}
\center
\caption{Measured values of $d_2^n$.}
\begin{tabular}{c c c c c}
$\left<Q^2\right>$  & Measured      & Low x         & Elastic      & Total $d_2^n$\\
(GeV$^2$/$c^2$)           &  ($\times$10$^{-5}$) &  ($\times$10$^{-5}$) & ($\times$10$^{-5}$) & ($\times$10$^{-5}$)\\
\hline

3.21 & -261.0 $\pm$ 79.0$_{\text{stat}}$ $\pm$ 48.0$_{\text{sys}}$ & 38.0 $\pm$ 58.0$_{\text{sys}}$ & -198.0 $\pm$ 32.0$_{\text{sys}}$  & -421.0 $\pm$ 79.0$_{\text{stat}}$ $\pm$ 82.0$_{\text{sys}}$\\

4.32 &  4.0   $\pm$ 83.0$_{\text{stat}}$ $\pm$ 37.0$_{\text{sys}}$ & 38.0 $\pm$ 58.0$_{\text{sys}}$ & -77.0  $\pm$  9.0$_{\text{sys}}$  & -35.0 $\pm$ 83.0$_{\text{stat}}$ $\pm$ 69.0$_{\text{sys}}$\\
\end{tabular}
\label{tbl:d2n}
\end{ruledtabular}
\end{table*}
\indent{}Primed with a  new value of $d_2^n$, we proceeded to determine $f_2^n$ and extract the average electric and magnetic color forces. The quantity $f_2^n$ was extracted following the analysis described in~\cite{Meziani:2004ne,posik_thesis}.~Our $f_2^n$ extraction used $a_2^n$ matrix elements evaluated from global analyses~\cite{deFlorian:2008mr,Bourrely:2001du,Bourrely:2007if,deFlorian:2005mw,Leader:2005ci,Gehrmann:1995ag}, which were found to be (4.3$\pm$12.1$)\times 10^{-4}$ and (0.6$\pm$11.3)$\times 10^{-4}$ at $\left<Q^2\right>$ = 3.21 and 4.32 GeV$^2$/$c^2$, respectively; our measured $d_2^n$ values; and the inclusion of the $\Gamma_1$ data from the JLab RSS experiment~\cite{Slifer:2008xu} and the most recent JLab E94-010 data~\cite{Slifer:2008re}.~The singlet axial charge, $\Delta \Sigma$, was determined from values of $\Gamma_1^n$  at $Q^2 \ge 5$ GeV$^2$/$c^2$ to be 0.375 $\pm$ 0.052, in excellent agreement with that found in~\cite{Accardi:2012qut}. We note that our extracted $f_2^n$ values are consistent with the value found in~\cite{Meziani:2004ne}.~A summary of our $f_2^n$ and average color force values, along with calculations from several models, are presented in Table~\ref{tbl:color_force}.\\
\begin{table*}[!]
\begin{ruledtabular}
\centering
\caption{Our results for $f_{2}^{n}$, $F_{E}^{n}$ and $F_{B}^{n}$ compared to model calculations. The value for $d_{2}^{n}$ is assumed to be zero in the instanton model calculation, as it is much smaller than $f_{2}^{n}$~\cite{Balla:1997hf}. Note that we have divided Eqs.~\ref{cf:f2d2_1} and \ref{cf:f2d2} by $\hbar c$ to obtain force units of MeV/fm.}
\begin{tabular}{c c l l l}
 Group                               & $Q^{2}$ (GeV$^{\text{2}}$/$c^2$) & $f_{2}^{n}\times10^{-3}$  & $F_{E}^{n}$ (MeV/fm)   & $F_{B}^{n}$ (MeV/fm)  \\
\hline
 E06-014                             &  3.21    &  43.57 $\pm$ 0.79$_{\text{stat}}$ $\pm$ 39.38$_{\text{sys}}$ & -26.17 $\pm$ 1.32$_{\text{stat}}$ $\pm$ 29.35$_{\text{sys}}$ &  44.99 $\pm$ 2.43$_{\text{stat}}$ $\pm$ 29.43$_{\text{sys}}$  \\
 E06-014                             &  4.32    &  39.80 $\pm$ 0.83$_{\text{stat}}$ $\pm$ 39.38$_{\text{sys}}$ & -29.12 $\pm$ 1.38$_{\text{stat}}$ $\pm$ 29.34$_{\text{sys}}$ &  30.68 $\pm$ 2.55$_{\text{stat}}$ $\pm$ 29.40$_{\text{sys}}$  \\
 instanton~\cite{Balla:1997hf,Lee:2001ug}         &  0.40      &  38.0                               & -30.41                        &  30.41                    \\
 QCD sum rule~\cite{Stein:1994zk,Stein:1995si}  &  1       & -13.0   $\pm$ 6.0                 &  54.25 $\pm$ 15.52            &  79.52 $\pm$ 30.06                  \\
 QCD sum rule~\cite{Balitsky:1989jb} &  1       &  10.0   $\pm$ 10.0                 &  29.73 $\pm$ 16.62            &  81.75 $\pm$ 30.64                  \\
\end{tabular}
\label{tbl:color_force}
\end{ruledtabular}
\end{table*}
\indent{}In summary, we have measured the DSA and unpolarized cross sections from a polarized $^3$He target, allowing for the precision measurement of the neutron $d_2$. We find that $d_2^n$ is small and negative at $\left<Q^2\right>$ = 3.21 and 4.32 GeV$^2$/$c^2$. We find that our results are consistent with the lattice QCD~\cite{Gockeler:2005vw}, bag model~\cite{Song:1996ea} and chiral soliton~\cite{Weigel:1996jh} predictions. We used our $d_2^n$ measurements to extract the twist-4 matrix element $f_2^n$ and performed a decomposition into neutron average electric and magnetic color forces. We note that our extracted $f_2^n$ is larger than our measured $d_2^n$, implying that the neutron electric and magnetic color forces are nearly equal in magnitude but opposite in sign.\\
\indent{}We would like to thank the JLab Hall A technical staff and Accelerator Division for their outstanding support, as well as M.~Burkardt, L.~P.~Gamberg, W.~Melnitchouk, A.~Metz, and J.~Soffer for their useful discussions. One of us (Z.-E.~M.) would like to particularly thank X.-D. Ji for his encouragement to propose and perform this measurement since 1995. This work was supported in part by DOE grants DE-FG02-87ER40315 and DE-FG02-94ER40844 (from Temple University). Jefferson Lab is operated by the Jefferson Science Associates, LLC, under DOE grant DE-AC05-060R23177.

\end{document}